\documentclass{emulateapj}
\usepackage{apjfonts}
\newcommand\lsim{\lesssim}
\newcommand\gsim{\gtrsim}
\newcommand\lya{Ly$\alpha$ }
\newcommand\lyb{Ly$\beta$ }
\newcommand\lyd{Ly$\delta$ }
\newcommand\lye{Ly$\epsilon$ }
\newcommand{\boldbeta}{\mbox{\boldmath$\beta$}}

\begin{document}

\title{Radiative Transfer Effect on Ultraviolet Pumping of the 21cm Line in
the High Redshift Universe}
\author{Leonid Chuzhoy\altaffilmark{1} and Zheng Zheng\altaffilmark{2,3}}

\altaffiltext{1}{McDonald Observatory and Department of Astronomy, The
University of Texas at Austin, RLM 16.206, Austin, TX 78712, USA;
chuzhoy@astro.as.utexas.edu}

\altaffiltext{2}{Institute for Advanced Study, Princeton, NJ, 08540, USA; zhengz@ias.edu}
\altaffiltext{3}{Hubble Fellow}

\begin{abstract}

During the epoch of reionization the 21cm signal is sensitive to the
scattering rate of the ultraviolet photons, redshifting across the
\lya resonance. Here we calculate the photon scattering rate profile
for a single ultraviolet source. After taking into account
previously neglected natural broadening of the resonance line, we
find that photons approach the resonance frequency and experience
most scatterings at a significantly smaller distance from the source
than naively expected $r=(\Delta\nu/\nu_0)(c/H)$, where $\Delta\nu=\nu-\nu_0$
is the initial frequency offset, and the discrepancy increases as 
the initial frequency offset decreases.
As a consequence, the scattering rate $P_\alpha(r)$ drops much faster
with increasing distance than the previously assumed $1/r^2$
profile. Near the source, ($r\lsim 1$ comoving Mpc), the scattering
rate of photons that redshift into the Ly$\alpha$ resonance
converges to $P_\alpha(r)\propto r^{-7/3}$. The scattering rate of
\lya photons produced by splitting of photons that redshift into a
higher resonance (Ly$\gamma$, Ly$\delta$, etc.) is only weakly
affected by the radiative transfer, while the sum of scattering
rates of \lya photons produced from all higher resonances also
converges to $P_\alpha(r)\propto r^{-7/3}$ near the source.  At
$15<z<35$, on scales of $\sim$0.01--20$h^{-1}$Mpc (comoving), the
total scattering rate of \lya photons from all Lyman resonances is
found to be higher by a factor of $\sim 1+0.3[(1+z)/20]^{2/3}$ than
obtained without full radiative transfer. Consequently, during the
early stage of reionization, the differential brightness of 21cm
signal against the cosmic microwave background is also boosted by a
similar factor.
\end{abstract}


\section{Introduction}

The next generation of radio telescopes (e.g., LOFAR, MWA, SKA, and
21CMA)\footnote{Information on these telescopes can be found at
http://www.lofar.org/, http://www.haystack.mit.edu/ast/arrays/mwa/,
http://www.skatelescope.org/, and http://cosmo.bao.ac.cn/project.html/,
respectively.}
promises to open a first observational window into the epoch preceding
the end of reionization at $z\gsim 6$. By measuring the redshifted 21cm
signal from neutral hydrogen, the new telescopes can provide us with
information on the history of reionization, the nature of the first
radiation source, the spectrum of the primordial density perturbation
field, the cosmological parameters, the physical properties of dark
matter particles, etc. \citep[e.g.,][]{MMR,CM,LZ,Nus,McQ,BL,CAS,SV}.

The 21cm signal can be observed either in absorption or emission
against the cosmic microwave background (CMB), when the hydrogen spin
temperature, $T_s$, is different from the CMB temperature,
$T_{\rm CMB}$. The former is defined by the relative populations of the
hyperfine states of hydrogen atoms
\begin{equation}
\frac{n_{\rm upper}}{n_{\rm lower}}=3e^{-h\nu_*/kT_s},
\end{equation}
where $\nu_*=1.4$ GHz is the frequency of hydrogen hyperfine
transition. In the high-density gas clouds, which so far have been
the only detected sources of the 21cm signal, the decoupling of
$T_s$ from $T_{\rm CMB}$  is done by collisions between atoms,
which induce direct transitions between the hyperfine states and
couple the spin temperature to the hydrogen kinetic temperature,
$T_k$. In the intergalactic medium (IGM), on the other hand, after
$z\sim 30$ the density becomes too low for collisional coupling to
be effective and the decoupling of  $T_s$ from $T_{\rm CMB}$
can be effectively done only by scatterings of Ly$\alpha$ photons,
which likewise couple $T_s$ to $T_k$ \citep{Wout}. The
spin temperature is therefore a weighted function of $T_{\rm CMB}$
and $T_k$  \citep{f8}
\begin{equation}
T_s=\frac{T_{\rm CMB}+y_\alpha T_k+y_cT_k}{1+y_\alpha+y_c},
\end{equation}
where $y_c$  is the collisional coupling constant (which we
neglect throughout the paper). The radiative coupling constant,
$y_\alpha$, is
\begin{equation}
\label{eqn:y}
y_\alpha=\frac{16\pi^2 T_*e^2 f_{12}J_0}{27 A_{10}T_k m_e c}S_\alpha,
\end{equation}
where $f_{12}=0.4162$ is the oscillator strength of the Ly$\alpha$
transition, $T_*=h\nu_*/k=0.068$ K, $A_{10}$ is the spontaneous
emission coefficient of the hyperfine transition and $J_0$ is the
intensity at Ly$\alpha$ resonance, when the backreaction on the
incident photons caused by resonant scattering is neglected. For the
unperturbed IGM the backreaction correction, $S_\alpha$, is
\citep{CS6}
\begin{equation}
\label{back}
S_\alpha=e^{-0.37(1+z)^{1/2}T_k^{-2/3}}\left(1+\frac{0.4}{T_k}\right)^{-1}.
\end{equation}

The Ly$\alpha$ photons can be produced in several ways, including
recombinations, line cooling, and collisional excitation of atoms by
non-thermal electrons produced by X-rays \citep{CAS,Chen07}.
However, in case the reionization epoch was dominated by stellar
ultraviolet (UV) sources (which at present is the most popular theory)
most Ly$\alpha$ photons in the neutral IGM originate as photons between
Ly$\alpha$ and Ly-limit frequencies. Due to expansion of the
Universe, a fraction of these photons (those emitted between
Ly$\alpha$ and Ly$\beta$) would gradually redshift into the
Ly$\alpha$ resonance. The rest would redshift into one of the higher
resonances (generally the one which is just below their initial
frequency). The scatterings of high resonance photons produce
electron excitations to the $np$ ($n>2$) states, which are shortly
followed by deexcitations and emission either of the original photon
(in case the cascade goes directly to the ground state,
$np\rightarrow 1s$) or of several lower energy photons (in case the
cascade goes via some intermediate state). Since between $\sim 12\%$
and $23\%$ of cascades follow the latter route, after just a few
scatterings most of the high resonance photons will be split.
Depending on whether the last cascade goes via the $2p$ or $2s$ the
resulting photons will or will not include a Ly$\alpha$ photon.

In this paper, we calculate the Ly$\alpha$ scattering rate $P_\alpha(r)$
as a function of distance $r$ from the UV radiation source.
Contrary to the naive expectation, the scattering rate does not
evolve as $1/r^2$.  While $P_\alpha(r)\propto 1/r^2$ is a good
approximation at low redshifts when hydrogen is mostly ionized, at
high redshifts where the Gunn-Peterson optical depth to the
Ly$\alpha$ photons is extremely large, we find that the scattering
rate profile becomes much steeper. Therefore, until the radiation
intensity reaches the saturation levels (i.e., $y_\alpha\gg T_{\rm
CMB}/T_k$), the fluctuations of the 21cm signal would be
significantly stronger than previously estimated.

The paper is organized as following. In \S~2, we describe our
calculation of radiative transfer for ``continuum'' Ly$\alpha$
photons (i.e., photons that gradually redshift into the Ly$\alpha$
resonance). In \S~3, we describe the radiative transfer of
``injected'' Ly$\alpha$ photons (i.e., photons produced by splitting
of high resonance photons). In \S~4, we study the total scattering
rate from ``continuum'' and ``injected'' \lya photons. In \S~5, we
summarize and discuss our results. Throughout the paper, we adopt a
spatially-flat $\Lambda$CDM cosmological model with matter density
$\Omega_m=0.25$ and baryon density $\Omega_b=0.044$ in units of the
critical density and a Hubble constant $H_0\equiv 100h=72 \, {\rm
km\, s^{-1}Mpc^{-1}}$, which is consistent with the constraints from
the {\it WMAP} data \citep{Spergel07}.

\section{Radiative transfer -- ``Continuum'' photons}

\begin{figure}[t]
\epsscale{1.2} \plotone{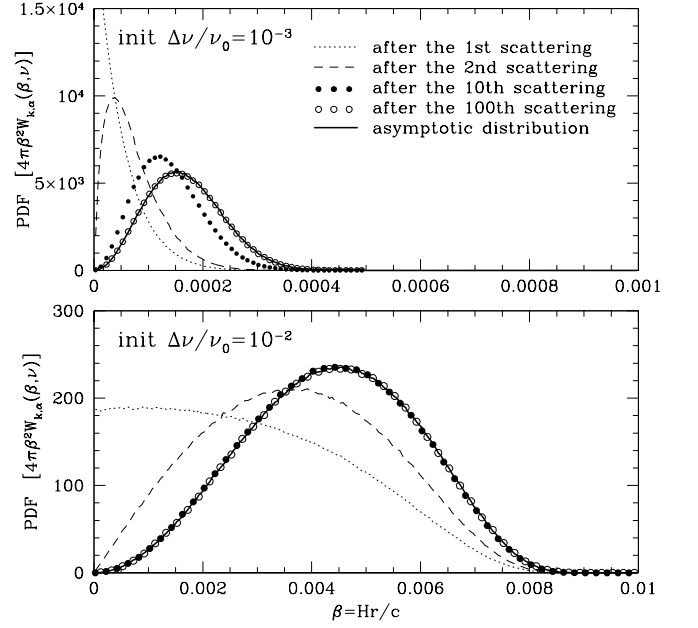} \caption[]{ \label{fig:con_nth}
Radial probability distribution function (PDF) of ``continuum'' \lya photons
around a central source at $z=25$, which is proportional to $r^2$ times
the \lya scattering rate $P_{\rm cont}(r)$.
The top and bottom panels are for \lya photons with
initial frequency offset $\Delta\nu/\nu_0$ of $10^{-3}$ and
$10^{-2}$, respectively. The dotted curve, dashed curve, filled circles,
and open circles are the distributions after the first, second,
10th and 100th scattering. The solid curve is the asymptotic
distribution before core scatterings happen, which is the same
profile of the scattering rate for photons with the given initial
frequency offset. If a Dirac $\delta_D$ function is adopted for the
scattering cross-section, the distribution would be a spike at the
right edge of each panel, with $Hr/c=\Delta\nu/\nu_0$. }
\end{figure}

The fate of the UV photon depends on the frequency at which it was
emitted. Most photons originally emitted between Ly$\alpha$ and
Ly$\beta$ frequencies will travel a large distance, up to
$(\nu_\beta/\nu_\alpha-1)c/H(z)\sim 300[(1+z)/26]^{-1/2}$ Mpc
(comoving), before being scattered by one of the hydrogen atoms. As
photon redshifts closer to the Ly$\alpha$ resonance its scatterings
become more frequent and the mean distance between subsequent
scatterings rapidly drops (see Figure \ref{fig:con_nth}). Therefore
almost all scatterings occur within a very small region\footnote{We find that
at $z\sim 25$ above 99\% of the scatterings occur within $\sim 1.5$ kpc
comoving.}. Thus to make an accurate estimate of the scattering
rate, it is in practice sufficient to count only scatterings
occurring within this region, where photons are very close to the
resonance and the number of scatterings is of the order of Gunn-Peterson
optical depth.  However, to derive the scattering rate profile,
$P_\alpha(r)$, one also needs to calculate the distance of this
region from the radiation source, which by contrast is determined
mainly by the first few scatterings before photons redshift into the
\lya resonance.

In general, photon scattering cross-section is the function of both
its frequency and the gas temperature. Neglecting hyperfine
splitting, the cross-section without thermal broadening can be written as
\begin{equation}
\label{eqn:cross}
\sigma(\nu)=\frac{\pi e^2}{m_e c} f \phi(\nu)
\end{equation}
with the line profile being a Lorentz function
\begin{equation}
\phi(\nu)=\frac{\gamma/4\pi^2}{(\nu-\nu_0)^2+(\gamma/4\pi)^2},
\end{equation}
where $\nu_0$ is the line-center frequency, $f$ is the oscillator
strength and $\gamma$ is the spontaneous decay rate ($\gamma/4\pi
\sim 10^{-8}\nu_0$ for \lya). Thermal broadening introduces a core
in the cross-section around the line center with a width of the
order of thermal velocity in units of light speed $c$, which is
about $10^{-6}$ for the unheated IGM in the redshift range we are
interested. In most calculations of the \lya scattering profile
around first sources, the line profile $\phi(\nu)$ is simplified to
a Dirac $\delta_D$ function and scatterings occur as a photon
redshifts into line-center frequency. We show in this paper that the
frequency dependence of the cross-section affects the scattering
rate profile.

Since in the high optical depth regime, the first scatterings happen
when photon frequency is still significantly above resonance (i.e.,
in the Lorentz wing), the distance between the source and the point
where photon enters into the resonance is nearly independent of gas
temperature. Hence the normalized scattering rate profile is also
independent of temperature [though not the total number of
scatterings, see eq.~(\ref{back})]. Since the line frequency offset
$\Delta\nu=\nu-\nu_0$ relevant for our study is much larger than the
quantum width and the thermal core width, the cross-section
[eq.~(\ref{eqn:cross})], being in the wing regime, can be well approximated
as
\begin{equation}
\sigma(\nu)=\sigma_c\left(\frac{\Delta\nu}{\nu_0}\right)^{-2},
\end{equation}
where $\sigma_c \propto f\gamma/\nu_0^2$ is formally the scattering
cross-section at twice the line-center frequency. We note that in this
study, the initial frequency of a photon is blueward of line center,
i.e., $\Delta\nu>0$. This is different from that in \citet{Loeb99}.
They investigate the brightness and spectral distributions of escaping
\lya radiation around sources before reionization, and in their case
photons start at a frequency slightly redward of the resonance, i.e.,
$\Delta\nu<0$.

\begin{figure}[t]
\epsscale{1.2}
\plotone{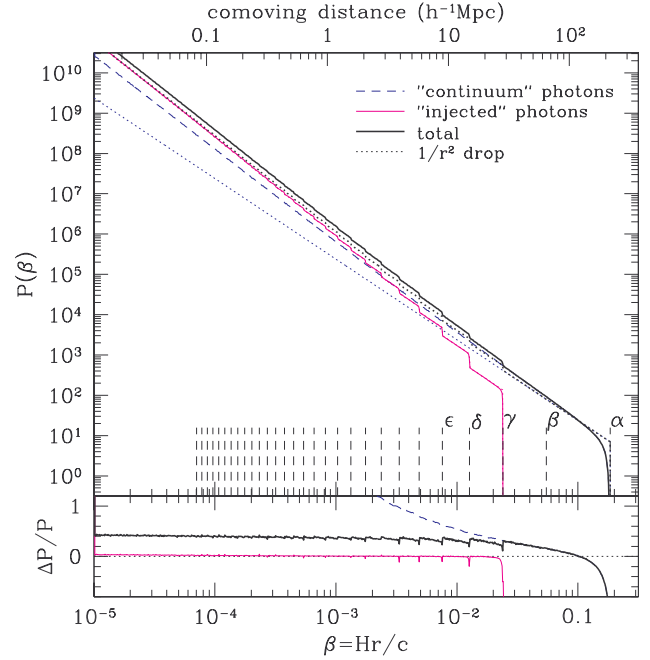}
\caption[]{
\label{fig:con_inj}
The radial profile of \lya scattering rate around the central source at $z=25$
as a sum of contributions from ``continuum'' and ``injected'' photons.
In the top panel, the blue (dashed), magenta (thin solid), and black (thick
solid) curves are those from ``continuum'' photons and
``injected'' photons and the total. The corresponding dotted curves are
obtained assuming the scattering cross-sections for Lyman lines to be Dirac
$\delta_D$ functions.  In the case of the ``injected''
photons, the dotted curve is almost on top of the magenta (thin solid) curve.
Dashed vertical lines mark the maximum distance reached
by Lyman series photons with the first five series labeled.
In the bottom panel, the fractional differences in the scattering rates with
respect to the case of $\delta_D$ cross-sections are plotted for the
``continuum'' photons (dashed blue), ``injected'' photons (thin magenta,
near 0), and the sum (thick black), respectively.
A flat UV spectrum is assumed (see the text).
}
\end{figure}

As a photon travels in the neutral medium with Hubble expansion,
its frequency redshifts. For a photon with initial frequency
offset $\Delta\nu$, the scattering optical depth to a distance $r$
is\footnote{We reduce the problem to a calculation for photons traveling
in a medium with Hubble expansion velocity field with fixed Hubble constant
at a given redshift. Strictly speaking, this approximation is only accurate
if the distance traveled is much less than the Hubble radius.
Such a condition is slightly broken for the largest scales in the
calculation, $\sim 20\%$ of the Hubble radius (see Fig.~\ref{fig:con_inj}),
which would introduce slight distortions in the scattering rate profiles on
these scales. However, for our main purpose, the scales in consideration are
generally much smaller and the above approximation is always valid.}
\begin{equation}
\label{eqn:tau} \tau=\int_0^r n_{\rm HI}\sigma(\nu-\nu Hr/c) dr =
\tau_c \beta \left[\frac{\Delta\nu}{\nu_0}
\left(\frac{\Delta\nu}{\nu_0}-\beta\right)\right]^{-1},
\end{equation}
where $n_{\rm HI}$ and $H$ are the neutral hydrogen number density and Hubble
constant at the redshift in consideration, $\beta=Hr/c$ and
$\tau_c=n_{\rm HI}\sigma_c c/H$. In what follows, we mainly adopt $\beta$ as
the distance variable, which proves to be convenient. At high redshifts,
the conversion from $\beta$ to comoving distance $d$ is simply
\begin{equation}
d = 1176 \beta \left(\frac{\Omega_m}{0.25}\right)^{-1/2}
    \left(\frac{1+z}{26}\right)^{-1/2} \,\, h^{-1}{\rm Mpc}.
\end{equation}
The parameter $\tau_c$ can be regarded as a redshift variable for
a given cosmology. Formally, it is the optical depth to the Hubble radius
seen by a photon with frequency twice the line-center frequency.
For \lya photons, we have
\begin{equation}
\label{eqn:tau_c}
\tau_c=0.0188 \left(\frac{h}{0.72}\right)
              \left(\frac{\Omega_b}{0.044}\right)
              \left(\frac{\Omega_m}{0.25}\right)^{-1/2}
              \left(\frac{1+z}{26}\right)^{3/2},
\end{equation}
where the mass fraction of helium is taken to be one quarter.

Equation~(\ref{eqn:tau}) can be inverted to find the distance traveled by
a photon of initial frequency offset $\Delta\nu/\nu_0$ for a given optical
depth $\tau$,
\begin{equation}
\label{eqn:distance}
\beta=\frac{\Delta\nu}{\nu_0}\left[1-\left(1+\frac{\Delta\nu}{\nu_0}\frac{\tau}{\tau_c}\right)^{-1}\right].
\end{equation}
For the solution to approach $\beta=\Delta\nu/\nu_0$, the one corresponding to
a Dirac $\delta_D$ cross-section, the frequency offset $\Delta\nu/\nu_0$ needs
to be large and/or the parameter $\tau_c$ needs to be small.
In the regime that $\Delta\nu/\nu_0\ll \tau_c$, $\beta$ is proportional to
$(\Delta\nu/\nu_0)^2$, which implies that on sufficiently small scales the
distance where most scatterings occur deviates from the expectation with
$\delta_D$ cross-section. The dependence on $\tau$ in
Equation~(\ref{eqn:distance}) also means that,
for a given initial frequency offset, the place where most scatterings occur 
is no longer at a single distance and instead it has a distribution.

The spherical symmetry and Hubble velocity field allow a simple
Monte Carlo calculation of the distribution of \lya photons from
continuum between \lya and \lyb as a function of distance from the
central source. The procedure is as follows:
\begin{itemize}
\item[i. ]{Start a photon from the center ($\beta_i=0$) with
           frequency offset $\Delta\nu/\nu_0$ drawn according to the
           slope of the UV spectrum of the central source.}
\item[ii.]{Draw a scattering optical depth $\tau$ according to the
           exponential distribution.}
\item[iii.]{Find the distance $\beta$ of traveling before scattering
            according to equation~(\ref{eqn:distance}).}
\item[iv.]{Determine the position of scattering,
           $\boldbeta_f=\boldbeta_i+\boldbeta$. That is,
           $\beta_f=\sqrt{\beta_i^2+\beta^2-2\beta_i\beta\cos\theta}$,
           where $\cos\theta$ is drawn uniformly between $-1$ and 1.}
\item[v.]{$\Delta\nu/\nu_0 \leftarrow \Delta\nu/\nu_0-\beta$ (i.e.,
           redshifted) and $\beta_i\leftarrow \beta_f$.}
\item[vi.]{Repeat ii.--v. until $\Delta\nu/\nu_0$ approaches the width of
           the thermal core (a few times $10^{-6}$ for situations here).
           Then start from i. again until the desired number of photons have
           been drawn.}
\end{itemize}
The cosmology and redshift are fully encoded in a single parameter $\tau_c$
[eq.~(\ref{eqn:tau_c})]. In the following discussions, we will assume
$\tau_c=0.0188$ for \lya photons, which corresponds to $z=25$.
In step iv., we simply assume the direction after scattering is isotropic.
We have tested that a more realistic angular distribution, such as a dipole
distribution, has little effect on the resultant spatial distribution of
\lya photons.

In Figure~\ref{fig:con_nth}, we show the spatial
distribution of \lya photons, $W_{k,\alpha}(r,\nu)$, after
$k$ scatterings for two initial frequency offsets $\Delta
\nu=\nu-\nu_\alpha$, based on the above Monte Carlo procedure. 
The distribution has been normalized so that 
$\int 4\pi\beta^2W_{k,\alpha}(\beta,\nu) d\beta=1$.
In each panel, the dotted curve, dashed curve, filled circles, and
solid circles are the distributions around the central source after
the 1st, 2nd, 10th and 100th scattering, respectively. As photons
redshift towards the line center, the scatterings become more
frequent and the distribution approaches an asymptotic one, 
$W_\alpha(r,\nu)$ (thick solid curve). Hence, in practice, to
derive the total scattering rate of \lya photons, it is sufficient
to perform the Monte Carlo simulation for the first few scatterings.

If the \lya cross-section is assumed to be a Dirac $\delta_D$
function, for an initial $\Delta\nu/\nu_0$ photon, all the
scatterings would happen at a distance $\beta=\Delta\nu/\nu_0$.
However, this is not the case if the frequency dependence of the
cross-section is taken into account, as shown in
Figure~\ref{fig:con_nth} (thick solid curves). The position where
most of the scatterings happen is broadly distributed at distances
smaller than $\beta=\Delta\nu/\nu_0$. Furthermore, the peak of the
distribution depends on the initial frequency, the smaller the
initial frequency offset $\Delta\nu/\nu_0$, the farther away of the
peak from the position expected from a $\delta_D$ function
cross-section. For example, as shown in Figure~\ref{fig:con_nth},
for $\Delta\nu/\nu_0=10^{-2}$, the peak of the distribution is at a
distance $\sim 45\%$ of that expected from a $\delta_D$ function
cross-section, while for $\Delta\nu/\nu_0=10^{-3}$ it is only $\sim
15\%$. As already mentioned, this dependence on initial frequency
can be understood by considering the first scattering with
equation~(\ref{eqn:distance}): for a given optical depth $\tau$,
$\beta$ becomes increasingly smaller than $\Delta\nu/\nu_0$ as
$\Delta\nu/\nu_0$ becomes smaller. Equation~(\ref{eqn:distance})
also shows that for smaller $\tau_c$, the scattering position would
become close to that expected from a $\delta_D$ function
cross-section, which we discuss more about in the next section.

To get the total scattering rate of the ``continuum'' Ly$\alpha$
photons for a general UV source, one needs to integrate over the
frequency range between Ly$\alpha$ and Ly$\beta$
\begin{equation}
\label{eqn:Pcont}
P_{\rm cont}(r)=
\tau_{\rm GP}S_\alpha
\int^{\nu_\beta}_{\nu_\alpha} W_\alpha(r,\nu) L_\nu d\nu,
\end{equation}
where $\tau_{\rm GP}$ is the Gunn-Peterson optical depth to \lya
photons, $S_\alpha$ is the backreaction correction factor
[eq.~(\ref{back})], and $L_\nu$ is the luminosity of the source in
terms of number of photons per unit frequency per unit time.

The above dependence of the scattering position on the initial
frequency implies that the radial profile of the scattering rate
would be steeper than the $1/r^2$ drop. With the Monte Carlo
technique,  we calculate the scattering rate profile from
``continuum'' photons, assuming a flat UV spectrum of the central
source (``flat'' here means equal number of photons per unit
frequency per unit time). In Figure~\ref{fig:con_inj}, the dashed
blue curve is the result from the Monte Carlo simulation, while the
dotted blue curve is the corresponding curve of the $1/r^2$ drop
with the same normalization of the UV spectrum of the central
source. The true profile is steeper than $1/r^2$ and the amplitude
can be an order of magnitude higher at small distances.

Close to the source the scattering rate scales as $P_{\rm
cont}(r)\propto r^{-7/3}$. This behavior can be explained by the
following argument. The mean free path $l$ of the photon emitted
with frequency offset $\Delta\nu$ (i.e., the average distance
between subsequent scatterings) scales as $l\propto (\Delta \nu)^2$
[eq.(\ref{eqn:distance}) in the limit of $\Delta\nu/\nu_0\ll
\tau_c\sim 10^{-2}$]. The change of frequency offset between two
subsequent scatterings is simply proportional to $l$. Therefore the
number of scatterings that would occur before the photon frequency
moves closer to resonance and its mean free path drops
significantly, scales as $N_{\rm sc}\propto \Delta\nu/l \propto (
\Delta\nu)^{-1}$. Since the photon direction changes almost at
random after each scattering, the total distance it travels until it
redshifts into the resonance scales as $r\propto \sqrt{N_{\rm
sc}}l\propto( \Delta\nu)^{3/2}$. Conversely, we can say that photons
reaching the resonance within distance $r$ from the source, are
emitted with frequency within $\Delta\nu\propto r^{2/3}$ from
Ly$\alpha$ resonance. Since the total number of photons within this
frequency range increases in direct proportion with $\Delta\nu$, the
photon scattering rate, which is proportional to the number of
photons reaching the resonance within some region divided by its
volume, scales as $P_{\rm cont}(r)\propto \Delta\nu/r^3 \propto
r^{-7/3}$.

\section{Radiative transfer -- ``Injected'' photons}

In addition to photons originally
emitted between Ly$\alpha$ and Ly$\beta$ frequencies, which redshift
into the Ly$\alpha$ resonance, Ly$\alpha$ photons are also produced
by splitting of photons originally emitted between Ly$\gamma$ and
Lyman-limit frequencies. These photons will be first scattered as
they redshift into the closest resonance.  In case, following their
absorption by a hydrogen atom, the excited electron cascades
directly to the ground state, the original photon would be
re-emitted. Alternatively, the electron can cascade via some
intermediate state, in which case the original photon is split into
several photons. Depending on the path of the cascade, the cascade
products may include the Ly$\alpha$ photon. The fraction of
Ly$\alpha$ photons made up by cascade of high resonances is
typically less than $\sim 15 \%$  of the total \citep{Hir,PF,CS7}.
However, as the distance traveled by photon before it redshifts into
the closest $np$ resonance is of order
$(\nu_{n+1}/\nu_n-1)c/H\propto n^{-3}$, the ``injected'' Ly$\alpha$
photons are produced within much smaller volume and within $\sim 10$
Mpc (comoving) from the source outnumber the ``continuum'' Ly$\alpha$
photons \citep{CAS}.

Since the ``injected'' Ly$\alpha$ photons are injected directly into
the resonance (hence their name), to derive their scattering rate
profile it is sufficient to follow the path of their high resonance
progenitors. The Monte Carlo simulation is similar to what we perform
for ``continuum'' photons, but we need to use cross-sections for higher
resonance lines [i.e., different $\tau_c$ and $\nu_0$ parameters in
eq.~(\ref{eqn:distance}) for
different lines] and record the probability of \lya production at each
scattering. In the calculation, we use the data in Table 1 of
\citet{PF} for the probability of decay from the $np$ state to the
ground $1s$ state and the probability of producing a \lya photon through
cascades from the $np$ level \citep[also see][]{Hir}. The scattering
cross-section for higher resonance Lyman lines are calculated from
oscillator strength and Einstein $A$ coefficient for $np\rightarrow 1s$
transition listed in \citet{Morton03} (and extrapolations are used for
higher $n$).

The total scattering rate of Ly$\alpha$ photons produced by cascade
from $np$ state with initial frequency $\nu=\nu_n+\Delta\nu$ is
\begin{equation}
P_{{\rm inj},n}(r)=\tau_{\rm GP} S_\alpha
\int^{\nu_{n+1}}_{\nu_{n}}\left[\sum_{k=1}^\infty f_{\alpha,n}
f_{{\rm des},n}(1-f_{{\rm des},n})^{k-1}W_{k,n}(r,\nu)\right] L_\nu d\nu,
\end{equation}
where $W_{k,n}(r,\nu)$ is the normalized spatial distribution for Ly$n$
photons that experience $k$ scatterings, $f_{{\rm des},n}$ is the probability
that the original Ly$n$ photon is destroyed, $f_{\alpha,n}$ is the chance that
the destruction of the Ly$n$ photon results in production of the
Ly$\alpha$ photon (i.e., that the electron cascade goes via $2p$ rather
than $2s$ state), and other symbols have the same meaning as in
equation~(\ref{eqn:Pcont}).
Since at each scattering there is a significant probability,
$0.12\lsim f_{{\rm des},n}\lsim 0.23$, that the original Ly$n$ photon is
destroyed, at large values of $k$, $(1-f_{{\rm des},n})^{k-1}$ approaches zero.
That is, the total distribution of ``injected'' \lya photons
can converge by considering those $np$ photons that experience only a small
number of scatterings.

\begin{figure}[t]
\epsscale{1.2} \plotone{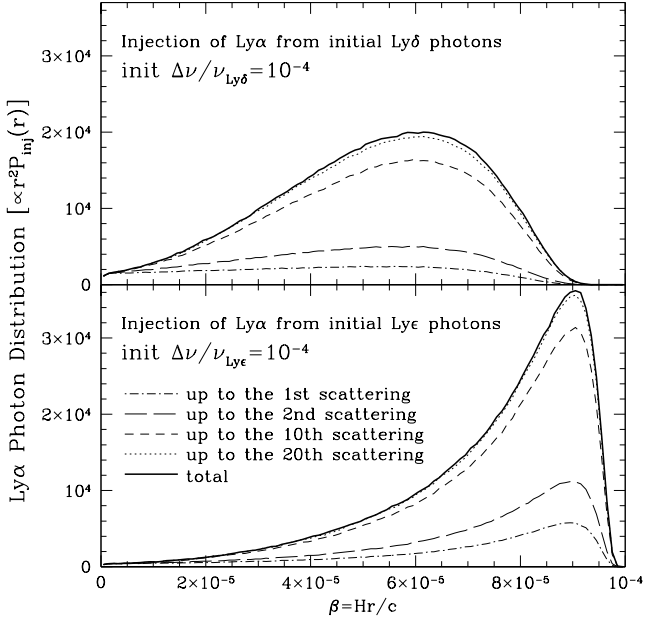} \caption[]{ \label{fig:inj}
Radial distribution of ``injected'' \lya photons around a central source at
$z=25$, which is proportional to $r^2$ times the \lya scattering rate
$P_{\rm inj}(r)$. Shown here are ``injected'' \lya photons from \lyd (top
panel) and \lye (bottom panel) resonances with initial frequency
offset $10^{-4}$. In each panel, the dot-dashed, long dashed, short
dashed, and dotted curves are contributions from Ly$n$
($n=\delta,\epsilon$) photons that experience no more than 1, 2, 10, and 20
scatterings before destroyed to produce \lya photons. The thick
solid curve is the total contribution from all Ly$n$ photons. The
distribution from a Dirac $\delta_D$ cross-section would be a spike
at the right edge of each panel, with $Hr/c=10^{-4}$. }
\end{figure}

Figure~\ref{fig:inj} illustrates the evolution of \lya photon
distribution for higher resonance photons (Ly$\delta$ and Ly$\epsilon$,
with initial frequency offset $10^{-4}$) that experience no more than $k$
scatterings ($k=$1, 2, 10, and 20) before destroyed. As expected, the
distribution approaches quickly to the total distribution (solid curve),
similar to the
``continuum'' case.
At the same initial frequency offset, higher
resonance photons lead to a peak position closer to that expected from
$\delta_D$ function cross-section. The reason is that a higher resonance
line has a smaller cross-section [thus a smaller $\tau_c$ parameter in
eq.~(\ref{eqn:distance})].

In contrast to ``continuum'' photons that typically reach the resonance
after multiple scatterings, a photon emitted with frequency blueward of
a higher Lyman resonance can produce a resonant \lya photon just
after a single scattering. If such a photon has mean free path $l\propto
(\Delta\nu)^2$ as the ``continuum'' photon case (see \S~2), the distance it
travels before producing a \lya photon is just
$r \propto l\propto (\Delta\nu)^2$ and one would expect the scattering
rate profile $P_{{\rm inj},n}(r) \propto \Delta\nu/r^3 \propto r^{-5/2}$.
However, this can only formally happen extremely close to the central
source [$\Delta\nu/\nu_n\ll \tau_c$, see eq.(\ref{eqn:distance})] because
of the fast drop of the corresponding $\tau_c$, and proximity
effect would take over at such scales.
Since both the oscillator strength $f$ and the spontaneous decay
rate $\gamma$ drops as $n^{-3}$ at large $n$, the characteristic scattering
optical depth $\tau_c$ in equation~(\ref{eqn:distance}) drops as $n^{-6}$.
Even for $n=3$, $\tau_c$ has dropped to a $10^{-5}$ level.
The low value of $\tau_c$ means that, at a
given initial frequency offset, the scattering rate distribution for
injected photons is almost the same as that expected from $\delta_D$
function cross-section with the peak position approaching
$\beta=\Delta \nu/\nu_n$.
For scales of interest (i.e., in the regime of $\Delta\nu/\nu_n > \tau_c$),
the mean free path $l\propto \Delta\nu$ instead of $(\Delta\nu)^2$ for
higher resonance photons, thus each individual $P_{{\rm inj},n}(r)$ follows
the $1/r^2$ profile.

The solid magenta curve in Figure~\ref{fig:con_inj} shows the \lya scattering
rate profile from injected photons for the same flat UV spectrum of the
central source as in the ``continuum'' case. Each of the steps in the curve
reflects the place that a higher resonance line starts to contribute to
produce injected \lya photons. At a given initial frequency offset, the
scattering rate of injected \lya photons is dominated by those produced by
higher level resonance lines. As mentioned above, the location of the injected
\lya photons from these resonance lines is close to the expectation of
$\delta_D$-form cross-section as the corresponding $\tau_c$ decreases
fast. Therefore, the true scattering rate profile is quite close to that
from $\delta_D$-form cross-section, which is a sum of a series of truncated
$1/r^2$ functions (dotted magenta curve in Figure~\ref{fig:con_inj}, almost on
top of the solid magenta curve). At a distance $\beta=10^{-5}$, the true
profile is only higher by $\sim$4\% in this particular case.

Although each individual $P_{{\rm inj},n}(r)$ follows a truncated
$1/r^2$ profile, the overall profile from summing over all
``injected'' photon contributions no longer follows a $1/r^2$ profile
because the truncation place for the individual component depends on
$n$. It can be shown that at small scales, the overall profile
$\sum_n P_{{\rm inj},n}(r) \propto r^{-7/3}$ \citep{CAS},
interestingly the same dependence as the ``continuum'' photon case
(see \S~2). To see this, we note that, with $r\propto
(\nu-\nu_n)/\nu_n$, the individual component $P_{{\rm inj},n}(r)$ is
simply $\propto L_\nu d\nu/(4\pi r^2 dr) \propto \nu_n/r^2$ for a
flat source spectrum $L_\nu=constant$ (in photon number per unit
frequency per unit time). The slope of the overall profile is then
$-2+d\ln(\sum_n \nu_n)/d\ln r$. Since $\nu_n \propto 1-1/n^2$, we
have $\sum_n\nu_n \propto n$ and $r\propto 1/n^3$ in the large $n$
limit. Therefore, the overall profile $\sum_n P_{{\rm inj},n}(r)
\propto r^{-7/3}$, which is what is seen in Figure~\ref{fig:con_inj}.

\section{The Total Scattering Rate}

The total scattering rate of Ly$\alpha$ photons is obtained after
summing over all Lyman series:
\begin{equation}
P_\alpha(r)=P_{\rm cont}(r)+\sum_{n=3}^\infty P_{{\rm inj},n}(r).
\end{equation}
The solid black curve in Figure~\ref{fig:con_inj} shows the profile
of the total scattering rate. For the case shown here ($z=25$),
``continuum'' (``injected'') photons dominate at distances to the center
greater (less) than $\beta\sim 2\times 10^{-3}$ ($\sim 3$Mpc
comoving). At smaller distance, even though the amplitude from
``continuum'' photons can be an order of magnitude higher than the naive
rate given by the $\delta_D$-form cross-section, the domination of
``injected'' photons reduces the difference in the total scattering
rate. As for the slope, the overall scattering rate is slightly
steeper than the sum of a series of truncated $1/r^2$ drops (dotted
black curve). To the first order, the overall scattering rate is
about 40\% higher than the naive calculation in a wide range of
distances (see the thick black curve in the lower panel).

The calculation of the scattering rate profile becomes much easier
by assuming a $\delta_D$-form scattering cross-section. From the
above example, we see that to the first order, the exact profile
that takes into account the frequency dependence of the
cross-section and the radiative transfer effect can be obtained by
applying a correction to that from the calculation with $\delta_D$
cross-section. This correction is almost a constant, increasing the
amplitude by a few tens of percent. Apparently, the correction
factor depends on redshift, larger at higher redshift because of an
increase in the $\tau_c$ parameter [eq.~(\ref{eqn:distance})]. We
perform simulations at a series of redshifts and find that the
fractional difference $f_c$ between the exact and the naive
calculations of the scattering rate can be approximated as a
constant at a given redshift,
\begin{equation}
f_c=0.381 \left(\frac{1+z}{26}\right)^{0.66}.
\end{equation}
This approximation underestimates (overestimates) the correction at
small (large) scales. For example, at $z=25$, $f_c$=44\% (32\%) at
$\beta=10^{-5}$ ($10^{-2}$). A more accurate fit that accounts for
such a tilt is
\begin{equation}
f_c=0.364 \left(\frac{1+z}{26}\right)^{0.71}
  +\left(0.0115\frac{z}{25}-0.0443\right)\log\frac{\beta}{10^{-3}},
\end{equation}
which works almost perfectly for $15<z<35$ and
$10^{-5}<\beta<10^{-2}$ (roughly corresponding to 0.01--20
$h^{-1}$Mpc comoving). The first term on the right hand side always
dominates and the second term accounts for the slight difference in
the slope of the two profiles. With this fitting formula, the
profile from the naive calculation can be corrected by multiplying
the scattering rate by a factor of $1+f_c$.

\begin{figure}[t]
\epsscale{1.2}
\plotone{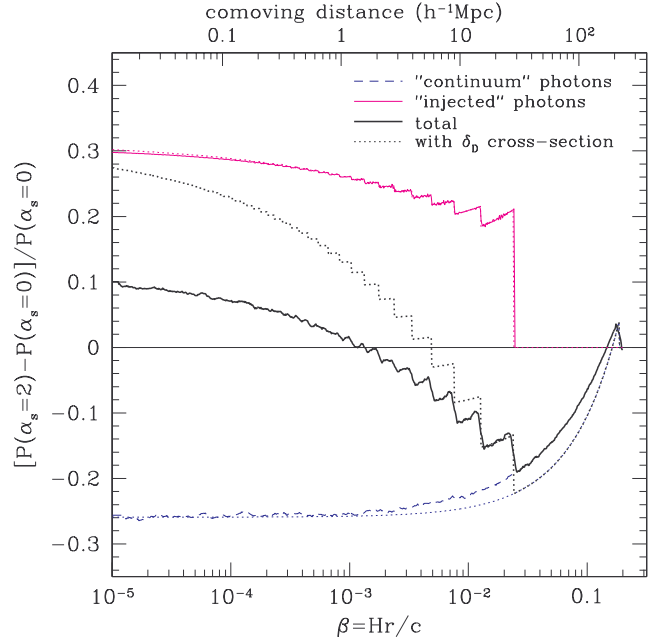}
\caption[]{
\label{fig:ratio}
Fractional differences in the scattering rate profiles for different
slopes of central source spectra at $z=25$. The luminosity of the
central source (number of photons per unit frequency per unit time) between
\lya and Lyman limit is assumed to follow $L_\nu\propto \nu^{\alpha_s-1}$
with $\alpha_s=2$ and 0, respectively. The spectra are normalized to
have the same number of photons between \lya and Lyman limit.
The dashed and thin solid curves are for rates from ``continuum'' and
``injected'' photons, respectively, and the thick solid curves is
for the total rate. Dotted curves are for the corresponding cases with
Dirac $\delta_D$ cross-sections.
}
\end{figure}

All the above examples assume a flat UV spectrum,
$L_\nu \propto \nu^{\alpha_s-1}$ with $\alpha_s=1$ between \lya and Lyman
limit, where $L_\nu$ is the luminosity in terms of number of photons per
unit frequency per unit time. The spectrum slope $\alpha_s$ depends on the
nature of the central source. For example, for Population III stars,
$\alpha_s=1.29$, while for Population II stars $\alpha_s=0.14$ \citep{BL}.
Given the narrow frequency range between \lya and Lyman limit, we do not
expect a strong dependence of the scattering rate profile on $\alpha_s$.
In the case of adopting the
$\delta_D$ cross-section, the scattering rate for an individual component 
is simply $\propto L_\nu d\nu/(4\pi\beta^2 d\beta)
\propto \nu_0^{\alpha_s}(1+\beta)^{\alpha_s-1}/\beta^2$, where the relation
$\beta=(\nu-\nu_0)/\nu_0$ is used. The departure from the $1/r^2$ drop
with respect to the flat spectrum is introduced through $\nu_0^{\alpha_s}$ and
$(1+\beta)^{\alpha_s-1}$ and both are small given the narrow frequency range
of Lyman series and $\beta\ll 1$.

We perform simulations for two cases with quite different spectral
slopes, $\alpha_s=2$ and $\alpha_s=0$. In Figure~\ref{fig:ratio}, we
plot ratios of scattering rates for the two cases. The central
source luminosity is normalized so that the numbers of photons emitted
between \lya and Lyman limit for the two cases are the same. That's
why the $\alpha_s=2$ case has a higher rate from ``injected''
photons (thin solid curve) but lower rate from ``continuum'' photons
(dashed curve). The ratios for the individual components are similar
to the $\delta_D$ cross-section results (dotted curves). However,
the steepening of the slope of the scattering rate for the
``continuum'' photons at small scales, which increases its
contribution to the total scattering rate, reduces the difference in
the total scattering rates for $\alpha_s=2$ and $\alpha_s=0$ (thick
solid curve). Even with the large difference in the spectral slope,
the difference in the overall scattering rates is only at a level
less than 10\% for $\beta<10^{-2}$. A comparison between cases using
spectral slopes of Population II ($\alpha_s=0.14$) and Population
III ($\alpha_s=1.29$) stars shows a pattern similar to what is seen
in Figure~\ref{fig:ratio} with the difference in the overall
scattering rates being at a level of 5\%.

\section{Summary and Discussion}

We investigate the effects of frequency dependence of scattering
cross-section and radiative transfer on the scattering rate
profile of Ly$\alpha$ photons around a central UV source, which
is relevant for the pumping of the 21cm line in the high redshift
universe.

Because of the frequency dependence of the scattering cross-section,
a photon between \lya and \lyb frequency (``continuum'' \lya photon)
experiences a small number of wing scatterings before it redshifts
to the core frequency and starts core scattering. Core scatterings,
which determine the scattering rate, happen at distances much less
than that expected from a $\delta_D$ scattering cross-section.  For
scatterings of higher resonance photons that produce ``injected''
\lya photons, the effect is weak owing to the fast drop in the
cross-section. We have shown that for a single UV source the
scattering rate profile of ``continuum'' Ly$\alpha$ photons is
significantly steeper than $1/r^2$ (previously expected from
$\delta_D$ scattering cross-section), approaching $r^{-7/3}$ near
the source. The scattering rate profile of ``injected'' \lya photons
from an individual high Lyman resonance closely follows a
truncated $1/r^2$ profile, while the overall profile from all higher
Lyman resonances coincidentally also approaches $r^{-7/3}$ near the
source. At $15<z<35$, the total scattering rate from ``continuum''
and ``injected'' photons is 30--50\% higher than the naive
calculation that adopts $\delta_D$ cross-sections, on scales of
$\sim$0.01--20$h^{-1}$Mpc (comoving). We also find that, when
radiative transfer effects are properly accounted for, the slope of
the scattering rate profile does not have a strong dependence on the
spectral slope of the central source.

In our calculations we have assumed that during the early stages of
reionization the gas temperature changes adiabatically. If instead,
the gas was significantly heated, the Doppler core for Ly$\alpha$
photons would be increased and their scattering rate profile would
be even steeper. However, since the temperature of neutral hydrogen
does not exceed $\sim 10^4$ K, the scales above $\sim 100$ kpc
(comoving) would be virtually unaffected.

As a consequence of the higher scattering rate with respect to the
naive calculation, the spatial fluctuations in the pumping radiation
field produced by multiple UV sources would be significantly higher
as well. During the early stages of reionization, when the UV
intensity is relatively low, the differential brightness of the 21cm
signal scales almost linearly with the scattering rate of the
Ly$\alpha$ photons and the corrections to the scattering rate
translate into similar corrections to the 21cm brightness. Such a
correction would increase the size of the expected \lya spheres
around first sources \citep[e.g.,][]{Cen06,Chen07}. Moreover, if the
radiative transfer effect is overlooked, the fluctuation power
spectrum estimated from 21cm observations would be skewed on scales
up to a few tens of Mpc, which would lead to inaccurate inference of
the matter fluctuation power spectrum (such as the amplitude and the 
running of the spectral index).
Therefore, until the epoch when the intensity of pumping radiation
reaches a saturation level (if such epoch in fact exists), the
correct interpretation of the 21cm signal requires taking into
account radiative transfer (mainly for ``continuum'' \lya photons),
as described in this paper.

\acknowledgments We thank Chris Hirata and Jordi Miralda-Escud\'e
for useful discussions. L.C. thanks the McDonald Observatory for the
W.J. McDonald Fellowship. Z.Z. acknowledges the support of NASA
through Hubble Fellowship grant HF-01181.01-A awarded by the Space
Telescope Science Institute, which is operated by the Association of
Universities for Research in Astronomy, Inc., for NASA, under
contract NAS 5-26555.

\end{document}